\documentclass[twocolumn,tighten]{aastex631}

\usepackage{aas_macros}
\usepackage[T1]{fontenc}
\usepackage{scrextend}
\usepackage{hyperref}
\usepackage{graphicx}
\usepackage{color}
\usepackage{natbib}
\usepackage{sidecap}
\usepackage{enumitem}
\usepackage{amsmath}
\usepackage{amssymb}
\usepackage{courier}
\usepackage{xfrac}

\hypersetup{
    colorlinks=true,
    linkcolor=blue,
    filecolor=magenta,      
    urlcolor=blue,
}

\setcitestyle{numbers}

\newcommand{\COBE}{\textsl{COBE}}

\newcommand{\WMAP}{\textsl{WMAP}}
\newcommand{\planck}{\textsl{Planck}}

\newcommand{\lcdm}{\ensuremath{\Lambda\mathrm{CDM}}}

\shorttitle{LAMBDA Overview}
\shortauthors{} 

\begin{document}

\title{Legacy Archive for Microwave Background Data Analysis (LAMBDA): An Overview}

\author{G.~E.~Addison}
\altaffiliation{Corresponding author, email gaddison@jhu.edu}
\affiliation{The William H. Miller III Department of Physics \& Astronomy, The Johns Hopkins University, 3400 N. Charles St., Baltimore, MD 21218-2686, USA}

\author{T.~M.~Essinger-Hileman}
\altaffiliation{LAMBDA Director}
\affiliation{NASA Goddard Space Flight Center, Greenbelt, MD, USA}

\author{M.~R.~Greason}
\affiliation{NASA Goddard Space Flight Center, Greenbelt, MD, USA}
\affiliation{ADNET Systems, Bethesda, MD, USA}

\author{T.~B.~Griswold}
\affiliation{NASA Goddard Space Flight Center, Greenbelt, MD, USA}
\affiliation{MORI Associates, Bethesda, MD, USA}

\author{T.~Jaffe}
\affiliation{NASA Goddard Space Flight Center, Greenbelt, MD, USA}

\author{N.~Miller}
\affiliation{The William H. Miller III Department of Physics \& Astronomy, The Johns Hopkins University, 3400 N. Charles St., Baltimore, MD 21218-2686, USA}
\affiliation{NASA Goddard Space Flight Center, Greenbelt, MD, USA}

\author{U.~Prasad}
\affiliation{NASA Goddard Space Flight Center, Greenbelt, MD, USA}
\affiliation{ADNET Systems, Bethesda, MD, USA}

\author{J.~L.~Weiland}
\affiliation{The William H. Miller III Department of Physics \& Astronomy, The Johns Hopkins University, 3400 N. Charles St., Baltimore, MD 21218-2686, USA}

\begin{abstract}

\noindent
This is an overview of the data products and other resources available through NASA's LAMBDA archive (\href{https://lambda.gsfc.nasa.gov/}{https://lambda.gsfc.nasa.gov/}). An up-to-date version of this document, along with code tools actively maintained and developed by LAMBDA staff, can be found on the LAMBDA GitHub page at \href{https://github.com/nasa-lambda/lambda_overview}{https://github.com/nasa-lambda/lambda\_overview}. New data products and other updates are announced on LAMBDA's X/Twitter account at \href{https://x.com/NASA_LAMBDA}{https://x.com/NASA\_LAMBDA}. If you have questions or suggestions relating to LAMBDA, or are interested in joining the LAMBDA Advisory Group, please contact us using the form here: \href{https://lambda.gsfc.nasa.gov/contact/contact.cfm}{https://lambda.gsfc.nasa.gov/contact/contact.cfm}.

\end{abstract}

\section{Introduction}

The Legacy Archive for Microwave Background Data Analysis (LAMBDA) was established in 2002 and in 2008 was made part of NASA's High Energy Astrophysics Science Archive Research Center (HEASARC, \href{https://heasarc.gsfc.nasa.gov/}{https://heasarc.gsfc.nasa.gov/}). LAMBDA provides a resource for cosmology researchers, hosting data products from cosmic microwave background (CMB) measurements, Galactic and extragalactic astrophysical microwave observations, and line intensity mapping (LIM) experiments, in addition to a range of software tools, and educational material.

\section{Data Products}

Below we provide embedded links to LAMBDA webpages, as well as references containing the full links. An overview of the full set of LAMBDA data holdings may be found \href{https://lambda.gsfc.nasa.gov/product/}{here} [\citenum{lambdadata}]. A table listing current, future and completed CMB anisotropy and spectrum experiments, with links to public data products, is provided \href{https://lambda.gsfc.nasa.gov/product/expt/}{here} [\citenum{lambdatable}]. An analogous table for LIM experiments is available \href{https://lambda.gsfc.nasa.gov/product/expt/lim_experiments.html}{here} [\citenum{lambdalimtable}].

LAMBDA is the primary data host for NASA's \href{https://lambda.gsfc.nasa.gov/product/cobe/}{COsmic Background Explorer (\COBE)} [\citenum{lambdacobe}] and \href{https://lambda.gsfc.nasa.gov/product/map/current/}{Wilkinson Microwave Anisotropy Probe (\WMAP)} [\citenum{lambdawmap}] CMB satellite missions. LAMBDA is also the primary NASA archive for \href{https://lambda.gsfc.nasa.gov/product/iras/}{Infrared Astronomical Satellite (IRAS)} [\citenum{lambdairas}] and \href{https://lambda.gsfc.nasa.gov/product/swas/}{Submillimeter Wave Astronomy Satellite (SWAS)} [\citenum{lambdaswas}] data. Additionally, we host data products from ground-based and suborbital collaborations, including:
\begin{itemize}
\item \href{https://lambda.gsfc.nasa.gov/product/abs/}{Atacama B-Mode Search (ABS)} [\citenum{abs2018}]
\item \href{https://lambda.gsfc.nasa.gov/product/act/}{Atacama Cosmology Telescope (ACT)}, including polarization-sensitive ACTPol, AdvACT data [\citenum{aiola/etal:2020}]
\item \href{https://lambda.gsfc.nasa.gov/product/bicepkeck/}{BICEP1, BICEP2, and Keck Array} [\citenum{bicep2keck:2021}] 
\item \href{https://lambda.gsfc.nasa.gov/product/class/}{Cosmology Large Angular Scale Surveyor (CLASS)} [\citenum{class2024}]
\item \href{https://lambda.gsfc.nasa.gov/product/polarbear/}{POLARBEAR} [\citenum{polarbear:2022}]
\item \href{https://lambda.gsfc.nasa.gov/product/quiet/}{Q/U Imaging ExperimenT (QUIET)} [\citenum{quiet:2012}]
\item \href{https://lambda.gsfc.nasa.gov/product/quijote/}{Q-U-I JOint Tenerife Experiment (QUIJOTE)} [\citenum{quijote:2023}]
\item \href{https://lambda.gsfc.nasa.gov/product/spider/}{SPIDER} [\citenum{spider:2022}] 
\item \href{https://lambda.gsfc.nasa.gov/product/spt/}{South Pole Telescope (SPT)}, including polarization-sensitive SPTpol, SPT-3G data [\citenum{spt3g:2023}] 
\end{itemize}
These data products include maps of CMB intensity and polarization fluctuations, weight or hit maps and noise simulations, as well as higher-level data products, such as likelihood code for fitting cosmological parameters.

LAMBDA also hosts ancillary data sets of interest to the CMB community, such as measurements of {\bf diffuse Galactic emission} for CMB or intensity mapping foreground studies. These data sets are available \href{https://lambda.gsfc.nasa.gov/product/foreground/fg_diffuse.cfm}{here} [\citenum{lambdafg}].

A collection of Galactic and extragalactic {\bf compact source catalogs} from radio, microwave, and infrared frequencies is provided \href{https://lambda.gsfc.nasa.gov/product/foreground/fg_comp_source.cfm}{here} [\citenum{lambdasources}].

{\bf Catalogs of clusters} detected using the Sunyaev Zel'dovich (SZ) effect from ACT,  \planck, and SPT, SZ angular power spectrum templates, and SZ simulations, are available \href{https://lambda.gsfc.nasa.gov/product/foreground/fg_sz_cluster.cfm}{here} [\citenum{lambdaclusters}].

\begin{figure}[t!]
    \centering
    \includegraphics[width=3.2in]{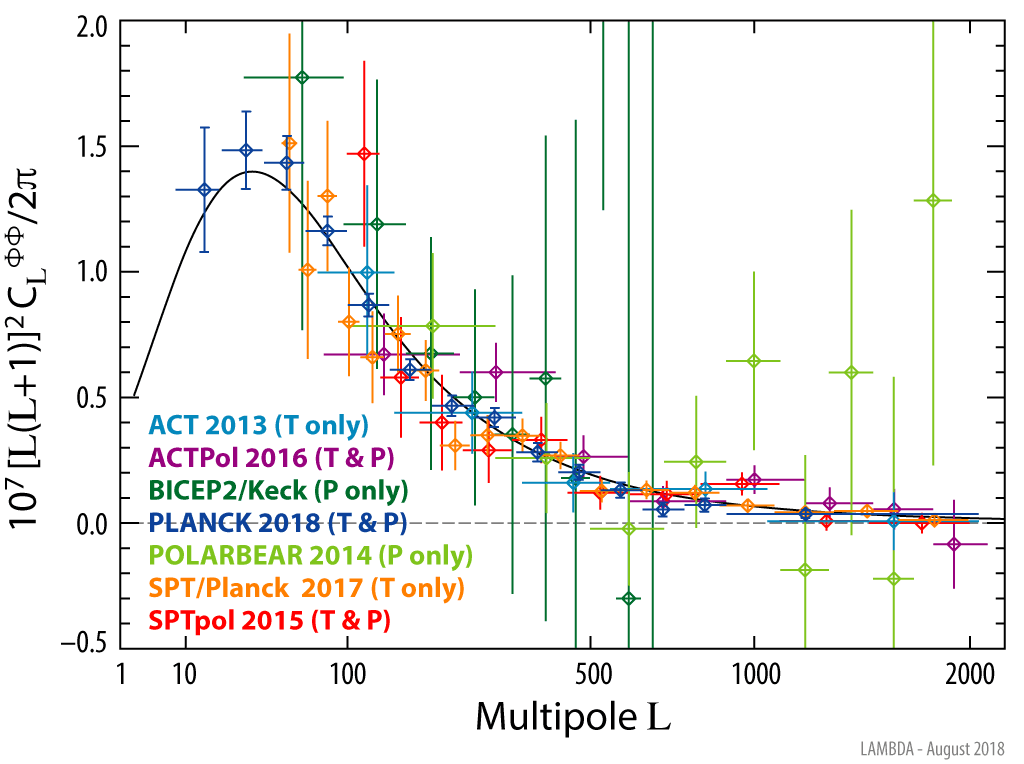}
    \caption{Compilation of CMB gravitational lensing potential power spectra from current experiments, estimated using temperature (`T') and polarization (`P') anisotropy data. This figure, and other figures comparing CMB temperature and polarization power spectra, including B-mode measurements, are available \href{https://lambda.gsfc.nasa.gov/graphics/}{here} [\citenum{lambdagraphics}], along with text files containing data points and error bars.}
    \label{fig:bandpower}
\end{figure}

Plots of CMB temperature, polarization, and gravitational lensing power spectra from various experiments, designed for use in presentations, are provided on the LAMBDA  \href{https://lambda.gsfc.nasa.gov/graphics/}{graphics page} [\citenum{lambdagraphicshome}]. A lensing bandpower plot is provided as an example in Figure~\ref{fig:bandpower}.

\section{Software Tools}

\begin{figure}[h!]
    \centering
    \includegraphics[width=3.2in]{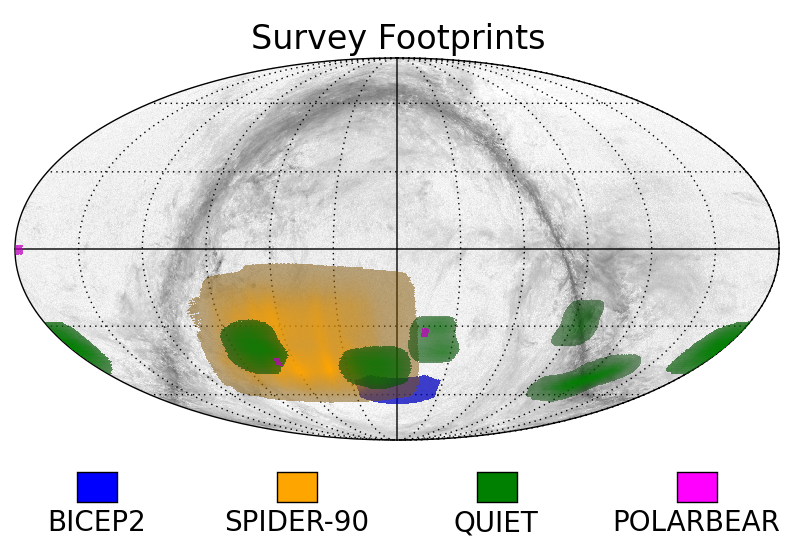}
    \caption{A comparison of the sky coverage of several CMB polarization surveys in equatorial coordinates, produced by the \texttt{cmb\_footprint} code. The grayscale background shows the Galactic polarized dust intensity measured by \planck.}
    \label{fig:footprint}
\end{figure}

Several tools are actively maintained or being developed by LAMBDA staff on GitHub \href{https://github.com/nasa-lambda}{here} [\citenum{lambdagithub}], including:
\begin{itemize}
\item \texttt{cmb\_footprint}, a tool to visualize footprints of different sky surveys. A sample set of CMB footprints from this tool is shown in Figure~\ref{fig:footprint}. An interactive version of the tool can be accessed on the main LAMBDA page \href{https://lambda.gsfc.nasa.gov/toolbox/footprint/aladin/}{here} [\citenum{lambdafootprint}].
\item \texttt{cmb\_analysis}, a set of analysis tools including code for calculating CMB power spectra and mode-mixing matrices from temperature and polarization maps with partial sky coverage
\item \texttt{cmbpol\_plotting}, data files and code for plotting and comparing CMB power spectra measurements from different experiments
\item \texttt{docker-lambda}, a Docker image file providing Python jupyter notebooks with many useful CMB utilities, including ACTPol and BICEP1 likelihood code examples
\end{itemize}

LAMBDA also hosts an interactive version of the \texttt{CAMB} [\citenum{lewis/etal:2000}] code for generating CMB power spectra  \href{https://lambda.gsfc.nasa.gov/toolbox/camb_online.html}{here} [\citenum{lambdacamb}], and provides links to several  \href{https://lambda.gsfc.nasa.gov/toolbox/converters.html}{unit conversion utilities} [\citenum{lambdaunit}], and \href{https://lambda.gsfc.nasa.gov/toolbox/calculators.html}{calculator utilities} [\citenum{lambdacalc}].

An extensive list of external tools for both CMB and broader astrophysical data analysis is provided with links \href{https://lambda.gsfc.nasa.gov/toolbox/}{here} [\citenum{lambdatoolbox}]. Examples include:
\begin{itemize}
    \item code for generating CMB and matter power spectra for a given set of cosmological parameters (e.g., \texttt{CAMB} [\citenum{lewis/etal:2000}], \texttt{CLASS} [\citenum{lesgourgues:2011}], \texttt{PICO} [\citenum{fendt/wandelt:2007a}])
    \item code for constraining cosmological parameter constraints from data using various Markov chain Monte Carlo methods (e.g., \texttt{CosmoMC} [\citenum{lewis/bridle:2002}], \texttt{MontePython} [\citenum{audren/etal:2013}], \texttt{Cobaya} [\citenum{cobaya:2021}])
    \item code made public by different experimental collaborations (ACT, BICEP2/Keck, SPT, etc.) for using their data in cosmological parameter fitting (e.g., new files to use with \texttt{CosmoMC})
    \item code for simulating aspects of the microwave sky, for example the Python Sky Model (\texttt{PySM} [\citenum{thorne/etal:2017}]), or the \texttt{Hammurabi} code for simulating polarized Galactic synchrotron emission [\citenum{waelkens/etal:2009}]
    \item tools relating to pixelization of the sky (\texttt{HEALPix}, \texttt{HEALPy} [\citenum{gorski/etal:2005}]), and calculating power spectra from pixelized sky maps (e.g., \texttt{PolSpice} [\citenum{chon/etal:2004}])
\end{itemize}

\section{Education}

LAMBDA hosts educational material, including several class and lab assignments provided by instructors, \href{https://lambda.gsfc.nasa.gov/education/}{here} [\citenum{lambdaedu}]. A graphical history of the \lcdm\ model, aimed at students, is provided \href{https://lambda.gsfc.nasa.gov/education/graphic_history/}{here} [\citenum{lambdagraphhist}]. This includes both theoretical material (e.g., description of components of the model and stages in evolution of the universe) and observational results, with graphical comparison of current and historical parameter constraints (e.g., age of the universe, optical depth, density of dark and baryonic matter, and Hubble constant).

\bibliographystyle{apj}

\begin{thebibliography}
\expandafter\ifx\csname natexlab\endcsname\relax\def\natexlab#1{#1}\fi

\bibitem{lambdadata}
\quad \url{https://lambda.gsfc.nasa.gov/product/}

\bibitem{lambdatable}
\quad \url{https://lambda.gsfc.nasa.gov/product/expt/}

\bibitem{lambdalimtable}
\quad \url{https://lambda.gsfc.nasa.gov/product/expt/lim_experiments.html}

\bibitem{lambdacobe}
\quad \url{https://lambda.gsfc.nasa.gov/product/cobe/}

\bibitem{lambdawmap}
\quad \url{https://lambda.gsfc.nasa.gov/product/map/current/}

\bibitem{lambdairas}
\quad \url{https://lambda.gsfc.nasa.gov/product/iras/}

\bibitem{lambdaswas}
\quad \url{https://lambda.gsfc.nasa.gov/product/swas/}

\bibitem{abs2018}
\quad {Akito}, K., {Appel}, J., {Essinger-Hileman}, T., {et~al.}, \jcap, 2018, 005; LAMBDA link \url{https://lambda.gsfc.nasa.gov/product/abs/}

\bibitem{aiola/etal:2020}
\quad {Aiola}, S., {Calabrese}, E., {Maurin}, L., {et~al.} 2020, \jcap, 2020, 047; LAMBDA link \url{https://lambda.gsfc.nasa.gov/product/act/}

\bibitem{bicep2keck:2021}
\quad {Ade}, P.~A.~R., {Ahmed}, Z., {Amiri}, M., {et~al.} 2021, \prl, 127, 151301; LAMBDA link \url{https://lambda.gsfc.nasa.gov/product/bicepkeck/}

\bibitem{class2024}
\quad {Eimer}, J.~R., {Li}, Y., {Brewer}, M.~K., {et~al.} 2024, \apj, 963, 92; LAMBDA link \url{https://lambda.gsfc.nasa.gov/product/class/}

\bibitem{polarbear:2022}
\quad {Adachi}, S., {Adkins}, T., {Aguilar Fa{\'u}ndez}, M.~A.~O., {et~al.} 2022,
  \apj, 931, 101; LAMBDA link \url{https://lambda.gsfc.nasa.gov/product/polarbear/}
  
\bibitem{quiet:2012}
\quad {QUIET Collaboration}. 2012, \apj, 760, 145; LAMBDA link \url{https://lambda.gsfc.nasa.gov/product/quiet/}

\bibitem{quijote:2023}
\quad {Rubi{\~n}o-Mart{\'\i}n}, J.~A., {Guidi}, F., {G{\'e}nova-Santos}, R.~T., {et~al.} 2023, \mnras, 519, 3; LAMBDA link \url{https://lambda.gsfc.nasa.gov/product/quijote/}

\bibitem{spider:2022}
\quad {Ade}, P.~A.~R., {Amiri}, M., {Benton}, S.~J., {et~al.} 2022, \apj, 927, 174; LAMBDA link \url{https://lambda.gsfc.nasa.gov/product/spider/}

\bibitem{spt3g:2023}
\quad Balkenhol, L., Dutcher, D., Spurio~Mancini, A., {et~al.} 2023, Phys. Rev. D,
  108, 023510; LAMBDA link \url{https://lambda.gsfc.nasa.gov/product/spt/}

\bibitem{lambdafg}
\quad \url{https://lambda.gsfc.nasa.gov/product/foreground/fg_diffuse.cfm}

\bibitem{lambdasources}
\quad \url{https://lambda.gsfc.nasa.gov/product/foreground/fg_comp_source.cfm}

\bibitem{lambdaclusters}
\quad \url{https://lambda.gsfc.nasa.gov/product/foreground/fg_sz_cluster.cfm}

\bibitem{lambdagraphics}
\quad \url{https://lambda.gsfc.nasa.gov/graphics/}

\bibitem{lambdagraphicshome}
\quad \url{https://lambda.gsfc.nasa.gov/education/lambda_graphics/index.html}

\bibitem{lambdagithub}
\quad \url{https://github.com/nasa-lambda}

\bibitem{lambdafootprint}
\quad \url{https://lambda.gsfc.nasa.gov/toolbox/footprint/aladin/}

\bibitem{lewis/etal:2000}
\quad {Lewis}, A., {Challinor}, A., \& {Lasenby}, A. 2000, \apj, 538, 473

\bibitem{lambdacamb}
\quad \url{https://lambda.gsfc.nasa.gov/toolbox/camb_online.html}

\bibitem{lambdaunit}
\quad \url{https://lambda.gsfc.nasa.gov/toolbox/converters.html}

\bibitem{lambdacalc}
\quad \url{https://lambda.gsfc.nasa.gov/toolbox/calculators.html}

\bibitem{lambdatoolbox}
\quad \url{https://lambda.gsfc.nasa.gov/toolbox/}

\bibitem[{{Lesgourgues}(2011)}]{lesgourgues:2011}
\quad {Lesgourgues}, J. 2011, ArXiv e-prints, arXiv:1104.2932


\bibitem[{{Fendt} \& {Wandelt}(2007)}]{fendt/wandelt:2007a}
\quad {Fendt}, W.~A., \& {Wandelt}, B.~D. 2007, \apj, 654, 2

\bibitem[{{Lewis} \& {Bridle}(2002)}]{lewis/bridle:2002}
\quad {Lewis}, A., \& {Bridle}, S. 2002, \prd, 66, 103511

\bibitem[{{Audren} {et~al.}(2013){Audren}, {Lesgourgues}, {Benabed}, \&
  {Prunet}}]{audren/etal:2013}
\quad {Audren}, B., {Lesgourgues}, J., {Benabed}, K., \& {Prunet}, S. 2013, Journal
  of Cosmology and Astro-Particle Physics, 2013, 001
  
\bibitem[{{Torrado} \& {Lewis}(2021)}]{cobaya:2021}
\quad {Torrado}, J., \& {Lewis}, A. 2021, \jcap, 2021, 057

\bibitem[{{Thorne} {et~al.}(2017){Thorne}, {Dunkley}, {Alonso}, \&
  {N{\ae}ss}}]{thorne/etal:2017}
\quad {Thorne}, B., {Dunkley}, J., {Alonso}, D., \& {N{\ae}ss}, S. 2017, \mnras, 469,
  2821

\bibitem{waelkens/etal:2009}
\quad {Waelkens}, A., {Jaffe}, T., {Reinecke}, M., {Kitaura}, F.~S., \& {En{\ss}lin},
  T.~A. 2009, \aap, 495, 697

\bibitem{gorski/etal:2005}
\quad {G{\'o}rski}, K.~M., {Hivon}, E., {Banday}, A.~J., {et~al.} 2005, \apj, 622,
  759

\bibitem{chon/etal:2004}
\quad {Chon}, G., {Challinor}, A., {Prunet}, S., {Hivon}, E., \& {Szapudi}, I. 2004,
  \mnras, 350, 914
  
\bibitem{lambdaedu}
\quad \url{https://lambda.gsfc.nasa.gov/education/}

\bibitem{lambdagraphhist}
\quad \url{https://lambda.gsfc.nasa.gov/education/graphic_history/}

\end{thebibliography}

\end{document}